\documentclass[dvips,11pt,a4paper]{article}

\usepackage{multibib}
\usepackage[T1]{fontenc}         
\usepackage[latin1]{inputenc}
\usepackage{graphicx}
\usepackage{amsmath}
\usepackage{bm}
\usepackage{tikz}
\usepackage{color}
\usepackage{helvet}
\usepackage{courier}
\usepackage{makeidx}
\usepackage{footmisc}
\usepackage{type1cm}         
\usepackage{textcomp}
\usepackage{subfig}
\usepackage{enumitem,xcolor}    
\usepackage{yhmath}

\DeclareCaptionLabelSeparator{colon}{. }
\usepackage[labelfont={bf},labelsep={colon}]{caption}

\DeclareMathAlphabet\mathbfcal{OMS}{cmsy}{b}{n}

\begin{document}

\begin{center}
\textcolor{blue}{ \Large  \bf  On primitive formulation in fluid mechanics and fluid-structure interaction  with constant piecewise properties in  velocity-potentials of acceleration } \\
\vspace{3.mm}
{\bf Jean-Paul Caltagirone$^1$, St{\'e}phane Vincent$^2$} \\
\vspace{3.mm}
{ \small  $^1$ Bordeaux INP - University of Bordeaux  \\
 I2M Institute, UMR CNRS $n^o$ 5295\\
  16 Avenue Pey-Berland, 33607 Pessac Cedex  \\
\textcolor{blue}{\texttt{ calta@ipb.fr }  } }  \\
\vspace{1.mm}
 { \small  $^2$ Universit{\'e} Paris-Est Marne-La-Vall{\'e}e,   \\
           Laboratoire Mod\'elisation et Simulation Multi Echelle (MSME),  UMR CNRS $n^o$ 8208, \\
          5 boulevard Descartes, 77454 Marne-la-Vall\'ee Cedex \\
  \textcolor{blue}{\texttt{stephane.vincent@u-pem.fr } }  }
\end{center}

\small
\textcolor{blue}{\bf Abstract}

Discrete mechanics makes it possible to formulate any problem of fluid mechanics or fluid-structure interaction in velocity and potentials of acceleration; the equation system consists of a single vector equation and potentials updates. The scalar potential of the acceleration represents the pressure stress and the vector potential is related to the rotational-shear stress.

The formulation of the equation of motion can be expressed in the form of a splitting which leads to an exact projection method; the application of the divergence operator to the discrete motion equation exhibits, without any approximation, a Poisson equation with constant coefficients on the scalar potential whatever the variations of the physical properties of the media. The {\it a posteriori} calculation of the pressure is made explicitly by introducing at this stage the local density.

Two first examples show the interest of the formulation presented on classical solutions of Navier-Stokes equations; similarly as other results obtained with this formulation, the convergence is of order two in space and time for all the quantities, velocity and potentials. This formulation is then applied to a two-phase flow driven by surface tension and partial wettability. The last case corresponds to a fluid-structure interaction problem for which an analytical solution exists.

\vspace{2.mm}
\textcolor{blue}{\bf Keywords}

Discrete Mechanics; Acceleration Conservation Principle; Hodge-Helmholtz Decomposition; Fluid-Structure Interaction; Navier-Stokes equation; Navier-Lam{\'e} equation 

\vspace{-3.mm}
\begin{verbatim}
___________________________________________________________________________
\end{verbatim}
\vspace{-2.mm}
NOTE: The final publication is available at link.springer.com
\vspace{3.mm}

J-P Caltagirone, S. Vincent, On primitive formulation in fluid mechanics and fluid-structure interaction  with constant piecewise properties in  velocity-potentials of acceleration, Acta Mechanica, 2020, \\
https://doi.org/10.1007/s00707-020-02630-w 
\vspace{-6.mm}
\begin{verbatim}
___________________________________________________________________________
\end{verbatim}

\vspace{-5.mm}

\textcolor{blue}{\section{Introduction} }

In fluid mechanics the dominant model is based on the Navier-Stokes equations where the principal variable is the velocity vector $ \mathbf V $ and the conservation of mass is ensured through a scalar, the pressure $ p $. In solid mechanics there are several formulations, in constraints or displacements; each one has its interest but the formulation in displacements, the Navier-Lam{\' e} equation, is more efficient to consider as it is easier to calculate the constraints if displacements are known than the opposite. The equations of mechanics are not unified, the Navier-Stokes equation requires to be accompanied by a conservation equation of mass to obtain the pressure whereas for the Navier-Lam{\' e} equation it is not required. A number of structural difficulties associated with the Navier-Stokes equation make it difficult to solve in a certain number of cases, in particular for two-phase flows.

The computation of the stresses and displacements in a solid in interaction with a moving fluid in complex situations imposes robust and precise numerical methodologies based on a reliable mathematical formulation. More and more work on these issues is using a monolithic approach to fluid-structure interaction.
However, the most delicate problem is posed for the Navier-Stokes equations which includes nonlinearities, which is not the case with the Navier-Lam{\' e} equation. The numerical treatment of Navier-Stokes equations in complex situations remains valid as is the theoretical order of the stability and convergence of the schemes used.

The formulation proposed here is based on a physical model different from that of Navier-Stokes. It is briefly recalled later but the details of the derivation of the discrete equations are developed in \cite{Cal19a}. The formulation based on the discrete mechanics allows access, at the same time, to the velocity $ \mathbf V $, compression stresses $\phi$ and shear $\bm \psi$; this formulation $( \mathbf V, \phi, \bm \psi )$ can be written as a projection type algorithm by a splitting or be used as it is.

Discrete geometric topologies are composed of a primal topology and a dual topology associated with specific variables where the scalars are located on the points of the primal geometry and the vectors on the edges of the latter. This vision is the generalization of the staggered Marker And Cell method on structured mesh. The polygonal or polyhedral meshes with any number of facets make it possible to pass over the usual 2D / 3D distinction for other methodologies.

The operators of the primal and dual geometric topologies defined here have obvious analogies with those of DEC (Discrete Exterior Calculus) method. This very elegant formalism based on Exterior Calculus takes its sources from Differential Geometry at the beginning of the last century; it has more recently been presented in a complete way, notably by Desbrun \cite{Des05} and Hirani \cite{Hirani2003}. Its applications are numerous in fluid mechanics, elasticity, electromagnetism, etc.

The main difficulties encountered in solving the Navier-Stokes equations with variable physical properties are developed first. Then the bases of the discrete formulation $ (\mathbf V, \phi, \bm \psi) $ from the concepts of discrete mechanics and objective advantages are highlighted. Examples of conventional problems of single fluid or two-phase flows are given to show that the solutions are in accordance with the results admitted in the literature. A case of fluid-structure interaction finally shows the merits of a unified approach.

\textcolor{blue}{\section{Formulations of Navier-Stokes equations} }

\textcolor{blue}{\subsection{Classical methodologies} }

The resolution of fluid flows is, in most cases, carried out using the Navier-Stokes equation and that of the solid interactions by the Navier-Lam{\' e} equation; these equations are not unified but many decoupled or monolithic techniques exist to solve fluid-structure interaction problems. Since the numerical difficulties encountered are mostly related to the Navier-Stokes equation, let's resume its most common form:
\begin{eqnarray}
\left\{
\begin{array}{llllll}
\displaystyle{ \rho \: \left( \frac{\partial \mathbf V}{\partial t} + \mathbf V \cdot \nabla  \mathbf V \right)  =  - \nabla p  +  \nabla \cdot \left( \mu \:  \left( \nabla \mathbf V + \nabla^t \mathbf V \right) \right)  } \\  \\
\displaystyle{ \frac{d \rho }{d t } + \rho \: \nabla \cdot {\mathbf V}  =  0  }
\end{array}
\right.
\label{NS}
\end{eqnarray}

In the general case of compressible flows with variable properties the application of differential operators, divergence or rotation on the equation (\ref{NS}) generate many difficulties. The introduction of a constitutive law and an energy conservation equation complicates the formulation for a compressible flow.

The case of incompressible flows where the conservation of the mass is replaced by the constraint $\nabla \cdot \mathbf V = 0$ removes a number of difficulties but also potentialities, for example to represent dilatable flows with variable density to less to adopt the Boussinesq approximation.
Even if the constraint is strictly verified, requires to adopt $\nabla \cdot \mathbf V = 0$, the application of the operators to the neighborhoods at the interfaces of the two-phase flows generate {\it artefact} which are due to the modeling of the Navier-Stokes equation.
If the viscosity remains a variable quantity in time and space, for example by the addition of a turbulent viscosity, the derivation of this quantity produces instabilities which are also due to the formulation of the viscous term of the equation (\ref{NS}).

The variations in density and viscosity lead to hypotheses, approximations and simplifications aimed at formulating the physical problem posed in an approximate manner. In particular, during the last decades, many works have been focused on the resolution of multimaterial problems and two-phase flows. Even if the methods prove to be efficient and precise, they do not palliate the errors induced by the adopted physical model.

All these difficulties have one and the same source, the notion of continuum. Is it legitimate to derive quantities that are physical properties of the media? Of course we cannot deny that there are variations of these properties within fluid flows or multimaterial problems but placing at the same level variables and physical properties has important consequences on the numerical methodologies used.

To ensure the constraint $ \nabla \cdot \mathbf V = 0 $ two large classes of methodologies exist, primitive variable pressure-velocity formulations $ (p, \mathbf V) $ and formulations of applying the curl operator to the Navier-Stokes equation including the one entitled curl-and-vector-potential formulation $ (\mathbf \Omega, \mathbf \Psi) $.

Among the methods using the primitive variables let us quote the method of artificial compressibility \cite{Pey83} where the pressure is reactualized starting from the divergence of the velocity computed by a step of prediction; it is an explicit method but does not require a condition at the limit on the pressure. Its implicit versions, augmented Lagrangian methods \cite{For82}, \cite{Vin04}, consist in integrating the constraint into the equation of motion as a Lagrangian; these are a little heavy from a numerical point of view but extremely effective. Direct methods are just as effective, they solve both velocity and pressure within a single algebraic system. These latter techniques, however, require robust and parallelizable solvers; if the case of two-dimensional problems can be treated by LU type direct solvers, the three-dimensional problems require iterative solvers, for example of preconditioned conjugate gradients.

The splitting of the Navier-Stokes equations historically derived by Chorin-Temam idea  and called fractional-step technique in 1968-1969 \cite{Cho67b} to which the projection methods are nowadays related, is very widely used for the resolution of fluid mechanics equations. J.L. Guermond's review \cite{Gue06} provides a mathematical and numerical point of view of these constant or variable density methods \cite{Gue09}. Many other variants analyze these methods from a mathematical point of view, \cite{Bel89}, \cite{Tru99}, \cite{Bro01}, \cite{Vre14}, and explicitly show convergence on numerical test cases.
Another point of view is developed in \cite{Cal99}, \cite{Ang12aa}, \cite{Cal15c} and \cite{Ang16b} to treat the problem of the update of pressure and that of the incompressibility of the flow which is considered exclusively kinematic. This method applies to highly variable density flows.

\textcolor{blue}{\subsection{Difficulties related to the term of inertia} }

All projection techniques involve, at one stage or another of the formulation, simplifications more or less justified; even if, from a practical point of view, they are minor and do not affect the representativeness of the solution obtained, they inhibit any chance of an exact projection.

The divergence of the incompressible Navier-Stokes equation given by Gresho ans Sani \cite{Gre87} with constant density is written:
\begin{eqnarray}
\displaystyle{  \nabla^2 p =  -  \nabla \cdot \left(  \mathbf V \cdot \nabla \mathbf V - \nu \: \nabla^2 \mathbf V  \right)  }
\label{divNS}
\end{eqnarray}
where $p$ is the pressure, $\nu$ the kinematic viscosity. If a source term $\mathbf f$  corresponding to gravitational or capillary effects is added to equation (\ref{NS}), the curl-free part of this term disappears thanks to the divergence operator.

The permutation of the divergence and gradient operators of the viscous term leads to a simplified form without this term; a part of the physics contained in the Navier-Stokes equation is thus lost. This confirms that the pressure is a simple Lagrangian that satisfies the constraint $ \nabla \cdot \mathbf V = 0 $. What remains anachronistic is that the pressure, a dynamic quantity, is used to satisfy a purely kinematic constraint. The point of view advanced here is that the quantity which makes it possible to satisfy the constraint of incompressibility must be purely kinematic. To show it let's examine the consequences of applying the divergence operator to the Navier-Stokes equation in more detail.

In the Navier-Stokes equations, the inertial term can be written according to the Lamb vector \cite{Lam93}, $\mathbfcal L = \nabla \times \mathbf V \times \mathbf V$. If we are interested in the divergence of this vector as Marmanis, \cite{Mar98}, Hamman \& al. \cite{Ham08}, Lindgren \cite{Lin12} or Kozachock \cite{Koz13} have been, it can be shown that $\nabla \cdot \mathbfcal L = \cdot \nabla \times \nabla \times \mathbf V - | \nabla \times \mathbf V |^2$. In this expression, the first term is called flexion while the second one is the local enstrophy. A rigid rotation movement is a local motion enstrophy as the rotational motion is at zero divergence but non-zero curl $\nabla \times \mathbf V = 2 \: \bm \omega$. In this section, the purpose is to examine the consequences of the expression of the divergence of the vector of acceleration $\bm \gamma$.

In the framework of continuum mechanics, the inertial term coming from the material derivative writes $\mathbf V \cdot \nabla \mathbf V$ or $- \mathbf V \times \nabla \times \mathbf V + \nabla (| \mathbf V |^2 / 2)$. By considering one of these two forms of the inertial terms, the divergence operator applied to them provides the same expression:  
\begin{eqnarray}
\displaystyle{  \nabla \cdot \left(  \frac{d \mathbf V}{dt} \right) =    \nabla \cdot \left( \frac{\partial \mathbf V }{\partial t} \right) + \mathbf V \cdot \nabla \left( \nabla \cdot \mathbf V \right) +   \left( \nabla \cdot \mathbf V \right)^2 - 2 \:  I_2  }
\label{divmecon1}
\end{eqnarray}
where the term $I_{2}$ is the second invariant of the velocity gradient, written in Cartesain coordinates:
\begin{eqnarray}
\hspace{-5mm}{\displaystyle {I_{2} = \left(\frac{\partial u}{\partial z}\:\frac{\partial w}{\partial x}+\frac{\partial v}{\partial z}\:\frac{\partial w}{\partial y}+\frac{\partial u}{\partial y}\:\frac{\partial v}{\partial x}-\frac{\partial u}{\partial x}\:\frac{\partial v}{\partial y}-\frac{\partial v}{\partial y}\:\frac{\partial w}{\partial z}-\frac{\partial w}{\partial z}\:\frac{\partial u}{\partial x} \right).}}
\label{I2}
\end{eqnarray}

If $u$ and $v$ are the components of the velocity in a surface then the term $- 2 \:  I_2$ can be represented by $\nabla u \wedge \nabla v $ where the gradient operator is applied in the planar surface $(x, y)$ and where $\wedge$ is the external product carried by the normal to this surface.

Is it possible to define a pseudo-vector $\bm I$ where components are associated to each surface defined $t_i$, $t_j$, the unit vectors on each axis for the three directions of space in Cartesian coordinates. Then  $I_{2}$  is defined as the scalar product  of $\bm I$   by the normal $\mathbf n = t_i \times t_j$ to each of the planar surfaces of the system.
The first three terms of the second member of the divergence of the material derivative (\ref{divmecon1}) are zero when the latter is equal to zero but the term $I_2$ is not. The reason why this term must be null is linked to vector conformity; $I_2$ is a scalar composed of the sum of the contributions defined by direction. The pseudo-vector $\bm I$ can thus be represented by the curl of a potential $\mathbf \Psi$ and the divergence of this term is necessarily zero $\nabla \cdot \nabla \times \mathbf \Psi = 0$. 
Although $ I_2$ is a scalar, each of the components of vector $\bm I$ has to be zero so that $\nabla \cdot \bm \gamma$ be zero too if the flow motion is incompressible $\nabla \cdot \mathbf V = 0$. The constraints $\bm I = 0$ are presented as compatibility conditions to be satisfied by component. They are similar to those used for defining displacements from constraints in solid mechanics. 
 Therefore residues from the Lamb vector will disturb the full projection of the velocity on a zero divergence field. The term $I_2$ is second order and numerically weak but the principle of an ``exact'' decomposition is compromised.

\textcolor{blue}{\section{Formulation in velocity-potentials} }

\textcolor{blue}{\subsection{Discrete Mechanics framework} }

The framework of discrete mechanics is composed of a geometric structure and a physical formulation based on intangible principles known since Galileo. The discrete geometric topology can be represented as the extension of the Marker And Cell method \cite{Har65} to unstructured staggered meshes.
\begin{figure}[!ht]
\begin{center}
\includegraphics[width=7.cm,height=4.6cm]{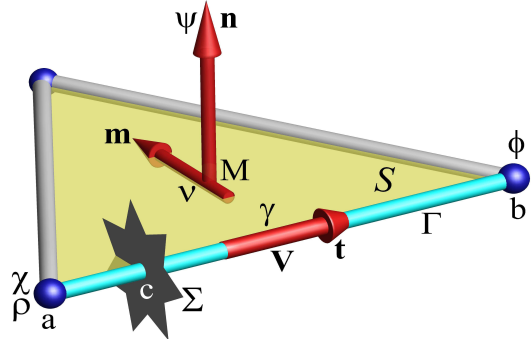}
\caption{\it Local frame of reference of discrete mechanics; each planar facet $ \mathcal S $ of normal $ \mathbf n $ whose centroid is at $ M $ is generated by rectilinear edges $ \Gamma $ of unit vector $ \mathbf t $, the trihedron $ (\mathbf m, \mathbf n, \mathbf t) $ being direct. A shock or contact discontinuity $ \Sigma $ can be located in $ c $ between the points of this primitive geometric topology. The scalar potential $ \phi $ is fixed on the points while the vector potential $ \bm \psi $ is oriented according to $ \mathbf n $; the $ \bm \gamma $ and $ \mathbf V $ components of the acceleration and velocity are constant on the $ \Gamma $ edge. The density $ \rho $ and the compressibility coefficient $ \chi $ are assigned to the point $ a $ or $ b $ and the kinematic viscosity is constant over the entire surface $ \mathcal S $. }
\label{discreet}
\end{center}
\end{figure}

The polygonal or polyhedral geometric structure is composed of elements such as that represented in the figure (\ref {discreet}). $\Gamma$ edges with unit vectors $ \mathbf t $ are connected by the points $ a $ and $ b $ of the primal geometric topology to form planar surfaces oriented along the normal $ \mathbf n $. The vectors, acceleration or velocity, or rather their components are defined on each edge where they are considered constant. The scalar potential $ \phi $ is defined at the point while the vector potential $ \bm \psi $ is carried by the normal $ \mathbf n $. The direct trihedron $ (\mathbf m, \mathbf n, \mathbf t) $ is the local coordinate system which serves as a basis for the representation of all physical quantities. A $\Sigma $ shock or contact discontinuity can be optionally located in $c$ on the $[a, b]$ edge.

In discrete mechanics the physical properties are not derivable, they are piecewise constant and assigned to different locations according to the modeled physical effect. The density $\rho$ is localized on the points but it is necessary to define a density $ \rho_v $ constant on the edge which depends on the position of $ c $; this is equal to $ \rho_v = \alpha \: \rho_b + (1 - \alpha) \: \rho_a$ with $ \alpha = [a, c] / [a, b] $. Likewise, a compressibility coefficient $ \chi $ is associated with each point of the primal topology. The kinematic viscosity $ \nu $ is itself defined at the center of the $ \mathcal S $ facet, but it is supposed to be constant on it.

The local frame of reference of figure (\ref {discreet}) allows to discard any global description and the use of an absolute or even Galilean reference. Uniform translation or rotation movements must be excluded from physical and mathematical modeling. All interactions between local frames of reference are cause-and-effect through common points or edges that will transmit information through propagation or diffusion from one element to another. The edge of length $d = [a, b]$ can be as small as necessary without ever being reduced to a point and the angles are preserved in a homothetic reduction. The notion of continuous medium is itself abandoned, which results in the loss of certain properties such as the derivation at a point.

We show \cite{Cal19a} that the Discrete Mechanics, the properties $ \nabla \times (\nabla \mathbf \phi) = 0 $ and $\nabla \cdot (\nabla \times \mathbf V) = 0$ are strictly satisfied for any continuous functions on polygonal or polyhedral non-structured topologies, regular or otherwise.

\textcolor{blue}{\subsection{Discrete law of motion} }

The physical modeling at the base of discrete mechanics is based on Galileo's intuitions: i) the Weak Equivalence Principle on the equivalence between gravitational and inertial effects (equality of the inertial and gravitational mass ) and ii) the relativity of velocity (Galilean reference frame). The fundamental principle of dynamics $\mathbf F = m_0 \: \bm \gamma$ where $m_0$ is the mass at rest and $\bm \gamma$ the proper acceleration of the body to which the force $\mathbf F$ is applied becomes $m_0 \: \bm g = m_0 \: \bm \gamma $ when applied to gravitation. The mass can thus be removed from this last expression to make it an equality between accelerations. The postulate of discrete mechanics consists in extending this form to all the other forces per unit of mass by designating by $\bm h$ the sum of the accelerations applied to the body. The equality $\bm \gamma = \bm h$ interprets the conservation of the accelerations: the proper acceleration of the particle or material medium is equal to the sum of the accelerations which are applied to it. The conservation of accelerations and energy replace those of mass and momentum of the continuum mechanics. 

The discrete mechanics has been the subject of some preliminary work \cite{Cal19a} in order to unify the mechanics of solids and incompressible or compressible fluids with or without shock waves. It reproduces, with the same formalism, the results deduced from the Navier-Lam{\'e} and Navier-Stokes equations, and those from other domains such as phase change transfer, porous media, and so on.

The law of motion in discrete mechanics is written in the form of a Hodge-Helmholtz decomposition of acceleration:
\begin{eqnarray}
\displaystyle{ \bm \gamma  = - \nabla \phi + \nabla \times \bm \psi  } 
\label{loidis}
\end{eqnarray}

Each component of the $ \bm \gamma $ acceleration can decompose a divergence free and an irrotational one, $ \bm \gamma_{\phi} = \nabla \phi $ and $ \bm \gamma_{\psi} = \nabla \times \bm \psi $; they both act without interaction on $ \Gamma $ because they are derived from orthogonal fields.
The equation of discrete motion is derived from the conservation equation of acceleration by expressing the deviations of the potentials $ \phi $ and $ \bm \psi $ as a function of the velocity $ \mathbf V $. These deviators are obtained on the basis of the physical analysis of the storage-destocking processes of compression and shear energies; the first is written as the divergence of velocity and the second as a dual curl velocity. The physical modeling of these terms is developed in \cite{Cal19a}.

The vectorial equation of the movement and its updates are written:
\begin{eqnarray}
\left\{
\begin{array}{llllll}
\displaystyle{ \bm \gamma = - \nabla \left( \phi^o - dt \: c_l^2 \: \nabla \cdot \mathbf V \right) + \nabla \times \left( \bm \psi^o - dt \: c_t^2 \: \nabla \times \mathbf V \right)   } \\ \\
\displaystyle{   \alpha_l \:\phi^o - dt \: c_l^2 \: \nabla \cdot \mathbf V \longmapsto \phi^o  } \\ \\
\displaystyle{  \alpha_t \: \bm \psi^o -  dt \: c_t^2 \: \nabla \times \mathbf V \longmapsto \bm \psi^o } \\ \\
\displaystyle{\mathbf V^o + \bm \: \gamma \: dt \longmapsto \mathbf V^o  }  \\  \\
\displaystyle{\mathbf U^o + \mathbf V^o \: dt \longmapsto \mathbf U^o  } 
\end{array}
\right.
\label{discrete}
\end{eqnarray}
where $\phi^o$ and $\bm \psi^o$ are the potentials of acceleration $ \bm \gamma $ defined at the instant $ t^o$; these potentials are updated at the moment $t$ to pass from one mechanical equilibrium to another over a period of time $dt$; the $\longmapsto$ symbol defined this upgrade. 

The quantities $c_l$ and $c_t$ are the longitudinal and transverse celerities. For fluids the definition of longitudinal velocity is equal to $c_l = \sqrt{1 / (\rho \: \chi_T)}$ where $\chi_T$ is the compressibility coefficient; the transverse celerity of a Newtonian fluid must be replaced by the kinematic viscosity, $dt \: c_t^2 = \mu / \rho$. For solids these parameters correspond to compressibility modules, $c_l = \sqrt{(\lambda + 2 \: \mu) / \rho}$, and shear $c_t = \sqrt{\mu / \rho}$.

 The factors $\alpha_i$ between zero and unity correspond to the attenuation of the longitudinal and transverse waves. The attenuation factor $\alpha_t$ of an elastic solid equal to unity while for a Newtonian fluid this factor is zero, the transverse waves are attenuated in general on very low time constants $(10^{-12} \: s)$ for fluids.

The variable is the component $\mathbf V$ of the velocity on each of the edges $\Gamma$. We observe that uniform velocity movements $ \mathbf V = Cte $ or uniform rotation $ \mathbf V = \omega \: r \: \mathbf e_{\theta} $ disappear from the outset of the vector equation of the system (\ref {discrete}). The velocity $ \mathbf V $ and the displacement $ \mathbf U $ are themselves upgraded from their value at the instant $t^o$.

Upgraded quantities $(\phi, \bm \psi)$ should be advected by the velocity field. This advection, independent of the equation (\ref{discrete}), can be realized in Eulerian by the resolution of a transport equation or in Lagrangian. Similarly, if density is not constant, for two-phase flows for example, it must be advected by an appropriate technique.

In the context of the choice of a fixed local reference, the proper acceleration $ \bm \gamma $ can be explained by highlighting the terms of inertia resulting from the discrete mechanics.
\begin{eqnarray}
\displaystyle{ \bm \gamma = \frac{\partial \mathbf V}{\partial t} + \nabla  \left( \frac{| \mathbf V |^2 }{2}  \right) - \nabla \times \left( \frac{| \mathbf V |^2 }{2} \: \mathbf n \right)  }
\label{inert}
\end{eqnarray}

The form of the inertial terms differs from that of the continuum mechanics, $ \mathbf V \cdot \nabla \mathbf V $ or $ \nabla (| \mathbf V |^2 / 2) - \mathbf V \times \nabla \times \mathbf V $ and also corresponds to a Hodge-Helmholtz decomposition. These terms may be partially implicited for better time discretization.

\textcolor{blue}{\subsection{Boundary conditions and orthogonality} }

Special attention has to be be paid concerning the boundary conditions to be applied in order to obtain an orthogonal Hodge-Helmholtz decomposition. This question is widely debated in textbooks about HHD and especially for its applications to the mechanics of fluids where the constraint of incompressibility is treated by projection methods. The need for orthogonality is discussed by Denaro \cite{Den03} and by Bhatia et al. \cite{Bha12}. Generally, in fluid mechanics, this is the decomposition of the velocity field that is sought in a free divergent part and an irrotational one; this is extended to the physical modeling of the movement of a fluid or a solid which leads to looking {\it a priori} to acceleration in the form of a HHD.

Boundary conditions can relate directly to velocity or its derivatives on the $\mathcal S$ surface of the $\Omega$ domain; in the context of the selected geometrical topology, the orthogonal fluxes at the boundaries are expressed only with the divergence of the velocity and the tangential stresses are represented only with the dual curl of the velocity. Boundary conditions are not linked as in a continuous environment and the concept of well-defined problem is not directly applicable.

Alternatively the boundary conditions can relate directly to the potentials and integrated into the discrete formulation in the form:
\begin{eqnarray}
\displaystyle{ \bm \gamma = - \nabla \left( \phi^o - dt \: c_l^2 \: \nabla \cdot \mathbf V + D \right) + 
 \nabla \times \left( \bm \psi^o - dt \: c_t^2 \: \nabla \times \mathbf V + \mathbf R \right) }
\label{condlim}
\end{eqnarray}
where $D = \nabla \cdot \mathbf V_n \: \mathcal S / dt$ is a flux per unit of time orthogonal to the domain and $\mathbf R = \nu \: \nabla \times \mathbf V_t$ a tangential viscous constraint. This equation is free from additional boundary conditions. From the $D$ and $\mathbf R$ border operators all types of boundary conditions can be imposed; they can also be used to impose a local source or well or a rotation of a subdomain within $\Omega$.

The legitimate question of orthogonality is discussed here by adopting the notations of  figure (\ref{discreet}) where $\mathbf n$ is the unit vector orthogonal to the surface  $\mathcal S$ of  the physical domain. Consider the two vectors $\mathbf f = \nabla \phi$ and $\mathbf g = \nabla \times \bm \psi$ and writing their inner product $<\mathbf f , \mathbf g>$ given by orthogonal decomposition theorems \cite{Hym99}, \cite{Ran19}:
\begin{eqnarray}
\left\{
\begin{array}{llllll}
\displaystyle{ \int_{\Omega} \nabla \phi \cdot  \nabla \times \bm \psi \: dv = \int_{\Omega}  \bm \psi \cdot \nabla \times  \left(  \nabla \phi \right)  \: dv + \int_{\mathcal S} \left(  \bm \psi \times \mathbf n \right)  \cdot  \nabla \phi \: ds  } \\  \\
\displaystyle{ \int_{\Omega} \nabla \phi \cdot \nabla \times \bm \psi \: dv = - \int_{\Omega} \: \phi \: \nabla \cdot \left( \nabla \times  \bm \psi \right) \: dv + \int_{\mathcal S}  \phi \: \left(  \nabla \times \bm \psi  \right) \: \cdot \mathbf n  \: ds  }
\end{array}
\right.
\label{ortho}
\end{eqnarray}

From the discrete point of view the equalities $\nabla_h \times \nabla_h \phi = 0$ and $\nabla_h \cdot \nabla_h \times \bm \psi = 0$ are strictly satisfied whatever the polygonal or polyhedral geometrical topologies in the whole physical domain; these properties are then verified on facet and each point of primal topology of figure (\ref{discreet}); $\bm \psi$ is collinear to $\mathbf n$ then $\bm \psi \times \mathbf n = 0$ and $\nabla \times \bm \psi$ is orthogonal to $\mathbf n$ so that $\nabla \times \bm \psi \: \cdot \mathbf n = 0$.
This result is verified for all facets $\mathcal S$ delimiting the polyhedron of the discrete domain wich defines $\Omega$. In both cases we have:
\begin{eqnarray}
\displaystyle{ \int_{\Omega} \nabla \phi \cdot \nabla \times \bm \psi \: dv = 0  }
\label{ortho2}
\end{eqnarray}

The HHD decomposition of the acceleration $\bm \gamma$ is orthogonal regardless of the BC for the proposed geometric description independently of the compressible or non-compressible nature of the flow.
\vspace{2.mm}

\textcolor{blue}{\subsection{An exact projection method} }

If the solutions of the equations of the discrete mechanics are the same as those of the Navier-Stokes equations with constant properties, it turns out that other differences appear when the dissipation term is extracted from these equations or if the divergence operator is applied to the acceleration. 
We have seen in continuum mechanics that the application of the divergence operator (\ref{divmecon1}) generated terms whose existence was due to the notion of continuum.

Consider now the divergence operator applied to the motion equation $\bm \gamma = - \nabla \phi + \nabla \times \bm \psi$, with leads to $\nabla \cdot \bm \gamma = - \nabla^2 \phi$.

The difference between continuum mechanics and discrete mechanics relates to the fact that $\nabla \cdot \bm \gamma = \nabla \cdot ( d \mathbf V / dt )$. In discrete mechanics, the particular derivative writes
\begin{eqnarray}
\displaystyle{ \frac{d \mathbf V}{dt} =    \frac{\partial \mathbf V }{\partial t} + \frac{1}{2} \:\nabla  \left( | \mathbf V |^2 \right)  - \frac{1}{2} \:  \nabla \times \left( | \mathbf V |^2 \: \mathbf n \right)  }
\label{divmecdis1}
\end{eqnarray}
and the divergence of it reads
\begin{eqnarray}
\displaystyle{ \nabla \cdot  \left( \frac{d \mathbf V}{dt} \right)  =    \nabla \cdot \left( \frac{\partial \mathbf V }{\partial t} \right)  + \frac{1}{2} \: \nabla \cdot \nabla \left( | \mathbf V |^2 \right)   }
\label{divmecdis2}
\end{eqnarray}

By introducing Bernoulli's potential $ \phi_B = \phi + | \mathbf V | ^ 2/2 $ and applying the divergence operator to the equation of motion it comes:
\begin{eqnarray}
\displaystyle{  \nabla \cdot \left( \frac{\partial \mathbf V }{\partial t} \right)   = - \nabla^2 \phi_B }
\label{divinert}
\end{eqnarray}

In order to express the partial derivative in terms of differences, $\mathbf V^{n + 1}$ is called  the zero divergence velocity at time $t^{n + 1}$ and its prediction $\widetilde{\mathbf V}$. The difference of the Bernoulli scalar potential between two times is denoted $\phi_B'= \phi_B^{n + 1} - \phi_B^n$.

The original projection algorithm based on discrete mechanics comes in the form:
\begin{eqnarray}
\left\{ 
\begin{array}{llllll}
\displaystyle{  \mathbf V^{n+1} - \widetilde{\mathbf V} =  - dt \: \nabla \phi_B' } \\ \\
\displaystyle{ dt \: \nabla^2 \phi_B' = \nabla \cdot \widetilde{\mathbf V}   }  \\  \\
\displaystyle{  \phi_B^{n+1} = \phi_B^n + \phi_B' } \\  \\
\displaystyle{  p_b^{n+1} = p_a^{n+1} - \int_{\Gamma} \: \rho_v \:  \nabla \phi_B^{n+1} \cdot \bm t \: dl } 
\end{array}
\right.
\label{algo}
\end{eqnarray}

As can be seen $ \phi_B '$ is extracted from the resolution of a Poisson equation with constant coefficients. The pressure, a dynamic quantity, uses the density $ \rho_v $ on the edge $ \Gamma $ and its computation is carried out completely explicitly directly on the mesh which can be 2D or 3D, structured or not. By choosing to fix the pressure at the first point $ a $ to an artibitrary value, we are gradually taking pressure on all the points of the primal mesh.

The jumps corresponding to discontinuities of shock or contact can be introduced directly into the vector equation in the form of a gradient or a primal curl. For example, the capillary pressure jump $ p_c $ can be implied in the form $ [[p_c]] = \nabla (\sigma \: \kappa) $ where $ \sigma $ is the surface tension per unit mass and $ \kappa $ the curvature on the $ \Sigma $ interface; if $ \rho_v = \alpha \: \rho_b + (1- \alpha) \: \rho_a $ and $ \alpha = [ac] / [ab] $ the pressure at $ b $ will be written: 
\begin{eqnarray}
\displaystyle{ p_b = p_a - \int_a^c \rho_a \: \nabla \phi_B \cdot \mathbf t \:  dl - \int_c^b \rho_b \: \nabla \phi_B \cdot \mathbf t \: dl  - \int_a^b   \nabla \left( \sigma \: \kappa \: \xi \right)\cdot \mathbf t  \: dl }
\label{kspc}
\end{eqnarray}
where the phase function $\xi$ assigned to each of the fluids will be equal to one or zero. The position $c$ of the $\Sigma$ interface on each edge will simply be known; in practice it can be calculated by the selected interface transport technique, Volume Of Fluid, Level-Set, Front-Tracking,  \cite{Osh88}, \cite{Sca99}, \cite{Pro07}, etc.

The problem of the residual terms that appears with the application of the divergence operator on the motion equation, in particular the viscous terms, no longer exists in discrete mechanics. Indeed, the viscosity is only present within the dual curl reminding that the properties discrete $ \nabla_h \times (\nabla_h \mathbf \phi) = 0 $ and $ \nabla_h \cdot (\nabla_h \times \mathbf V) = 0 $ are strictly checked for any type of elementary topology.

The term ''exact'' used in this section refers to a projection method directly derived from the divergence of the equation of motion. It turns out that the numerical differences with conventional projection methods are low with constant physical properties but can generate instabilities when density ratios or viscosities are high.

\textcolor{blue}{\subsection{Discrete formulation properties} }

In continuum mechanics pressure is the scalar potential of acceleration; it is also the case of the scalar potential $ \phi $ which replaces the quantity $ p / \rho_v $ including for flows with variable density. The density $ \rho_v $ on the edge $ \Gamma $ being constant the product $ (\rho \: \mathbf V) $ is the momentum, also constant in pieces.

The classical rotational-potential vector formulation reveals the rotational vector $\mathbf \Omega = \nabla \times \mathbf V$ and the vector potential of velocity, $\mathbf V = \nabla \times \mathbf \Psi$.
But in discrete mechanics $\bm \psi$ is neither the curl of $\mathbf V$ nor the vector potential of $ \mathbf V$, this is the vector potential of the acceleration. This quantity is not a kinematic operator, it represents a physical reality $\bm \psi = \nu \: \nabla \times \mathbf V$ is the instantaneous spin-shear stress.

The variables $ (\mathbf V, \phi, \bm \psi) $ correspond to a real formulation in primitive variables; the velocity is calculated from equation (\ref {discrete}) and the potentials are explicitly upgraded according to only a step.

The advantages of the formulation $ (\mathbf V, \phi, \bm \psi) $ can be synthesized into a list:
\begin{itemize} [label =  {\textbullet}]
\item the treatment of compressible or incompressible fluid flows, with or without shock waves, the computation of the stresses and displacements in the solids and the fluid-structure interaction can be accomplished with one and the same system of equations;
\item velocity, compression and shear stresses are obtained in the same step;
\item the method used on many cases of fluid or two-phase flows is particularly robust and precise; it is second order in space and time on velocity and potentials if temporal discretization is carried out with Gear type schemes;
\item it is applicable for non-constant physical properties, density and highly variable viscosities; in particular the fluid-solid interaction is produced without disturbance to the interfaces;
\item the interfaces are taken into account without any interpolation, in a sharp way; the primal mesh is plated on the underlying materials or fluids;
\item all quantities are expressed only with two basic units, a length and a time;
\item the formulation excludes tensors, so there is no vector gradient to calculate; it is of the ready-to-use type, the discrete operators correspond to the spatial discretization;
\item the integration of jumps of the physical properties but also of the variables is realized directly in the vector equation of the movement in the form of gradient or dual curl;
\item the declination of this formulation into a splitting provides the opportunity for an exact projection method by applying the divergence operator to acceleration.
\end{itemize}

These concrete advantages come directly from a revision of the equations of mechanics and the more fundamental one of the notion of continuous medium. If the one-point derivation of the variables is natural, that of the physical properties introduces many difficulties in the formulation of the equations of physics. The discrete medium in no way diminishes their scope and generalization.

\textcolor{blue}{\section{Examples} }

Numerous examples have shown the validity of the system (\ref{discrete}) in fluid mechanics. For example, compressible, incompressible, non-isothermal or two-phase flows can be approximated with DM to recover the standard well known results, in particular the classical analytical solutions of Poiseuille or Couette flows.  Reference cases, such as the lid-driven cavity, the backward-facing step or the flow around a cylinder, make it possible to show that the DM model converges to order two in space and time for both velocity and pressure. The flows associated with heat and mass transfers including multi-components are reproduced in a similar way. More complex problems of shock waves, like the Sod tube, phase changes, boiling and condensation \cite{Ami14} are treated in a coherent way by integrating discontinuities within the equations of motion. Two-phase flows with capillary effects, surface tension or partial wetting, are particularly well suited to the discrete mechanics model \cite{Cal15c}.

In the present form, the system (\ref{discrete}) is relatively close to the Navier-Lam{\'e} equation associated with the study of stresses and displacements in solids. It differs, however, on several points: more particularly, the discrete formulation is established in velocity and the displacement is only an accumulation of $\mathbf V \: dt$, as the velocity is itself the elevation of $\bm \gamma \: dt $. Numerous examples of simple solicitations make it possible to find the solutions of the Navier-Lam{\'e} equation. More complex 2D and 3D problems on monolithic fluid-structure interactions have already made it possible to validate the proposed formulation \cite{Bor14}, \cite{Bor16}. The vision of a continuous memory medium makes it possible to treat the problems of large deformations and large displacements in a formulation where pressure stress and shear are obtained synchronously without compatibility conditions. Given the original dissociation between compression effects and rotation, the material frame-indifference introduced by Truesdell \cite{Tru74} is satisfied naturally. The complex constitutive laws can be treated without difficulty and only the physical parameters written in the equation of motion must be known.

\textcolor{blue}{\subsection{The Green-Taylor vortex} }

The Green-Taylor vortex is a synthetic analytical solution of the Navier-Stokes equations that corresponds to a case of incompressible unsteady flow. This case is often used to compute convergence orders in time and space of numerical methods. It is considered here to show that the discrete formulation makes it possible to find the second-order analytical solution in time and space. More particularly, it should enable us to understand the role played by the different terms of the discrete motion equation and also to characterize how they combine to satisfy operator properties identically.

The equation of the unsteady movement reads:
\begin{eqnarray}
\hspace{-10.mm}
\displaystyle{   \frac{\partial \mathbf V}{\partial t} + \nabla \phi_i - \nabla \times \bm \psi_i   =  - \nabla \left(  \phi^o  - r \: \nabla \cdot \mathbf V \right) - \nabla \times \left(  \nu \: \nabla \times \mathbf V  \right) + \mathbf S_{\mu} + \mathbf S_i }
\label{green}
\end{eqnarray}

 The quantities $\mathbf S_i = \partial \mathbf V / \partial t$ and $\mathbf S_{\mu}$ are suitable source terms that lead to a stationary solution. Parameter $r$ makes it possible to maintain, at any instant, the divergence of the velocity below $1 / r$; this can be considered as zero throughout the simulation. Since the medium is a Newtonian fluid, the accumulation of shear-rotation constraints is zero $\bm \psi^o = 0$ with the instantaneous potential being equal to $\bm \psi = - \nu \: \nabla \times \mathbf V$. In fact, the dual curl of the vector potential is equal to source term $\mathbf S_{\mu}$ for this form of the equation.
For $(x,y) \in [-0.5, 0.5]$, the solution is written:
\begin{eqnarray}
\hspace{-5.mm}
\left\{  
\begin{array}{llllll}
\displaystyle{ \phi =  \frac{V_0^2}{2} \: \left( cos(\pi x)^2 + cos(\pi y)^2 \right) \: \left( 1 - \exp(-\pi \: t) \right)^2 } \\ \\
\displaystyle{ u = - V_0 \: cos(\pi\: x) \:\: sin(\pi \:y)  \: \left( 1 - exp\left( - \pi \: V_0 \: t \right) \right)  }  \\  \\
\displaystyle{ v = \:\:\: V_0 \: sin(\pi\: x) \:\: cos(\pi \:y) \: \left( 1 - exp\left( - \pi \: V_0 \: t \right) \right)  }
\end{array}
\right.
\label{taylor}
\end{eqnarray}
with $\mathbf V = u(x, y) \: \mathbf e_x + v(x, y) \: \mathbf e_y$ and $\bm \psi = - \nu \: \nabla \times \mathbf V$.
 
The source term calculated from solution (\ref{taylor}) leads to a separation between inertial and viscous terms:
\begin{eqnarray}
\left\{  
\begin{array}{llllll}
\displaystyle{   \frac{\partial \mathbf V}{\partial t} = \mathbf S_i }  \\  \\
\displaystyle{   \nabla \times \left(\bm \psi^o - \nu \: \nabla \times \mathbf V \right)  = 0 }  \\  \\
\displaystyle{ - \nabla \times  \bm \psi_i  = - \nabla \phi_B  }
\end{array}
\right.
\label{green2}
\end{eqnarray}

The potential of Bernoulli $\phi_B = \phi + \phi_i$, where $\phi_i$ the inertial acceleration, is the equivalent of the Bernoulli pressure. We find that $\nabla \times \bm \psi^o = \mathbf S_{\mu}$. Adding this source term into the motion equation amounts to imposing a vector potential $\bm \psi^o$ depending on time. In discrete mechanics, equilibrium is not required by component and only the addition of accelerations on the edge $\Gamma$ makes it possible to translate the mechanical equilibrium. Here, the solution of the problem $(\mathbf V, \phi, \bm \psi) $ is obtained directly by the resolution of the vector equation (\ref{green}).

\begin{figure}[!ht]
\begin{center}
\includegraphics[width=5.6cm,height=4.6cm]{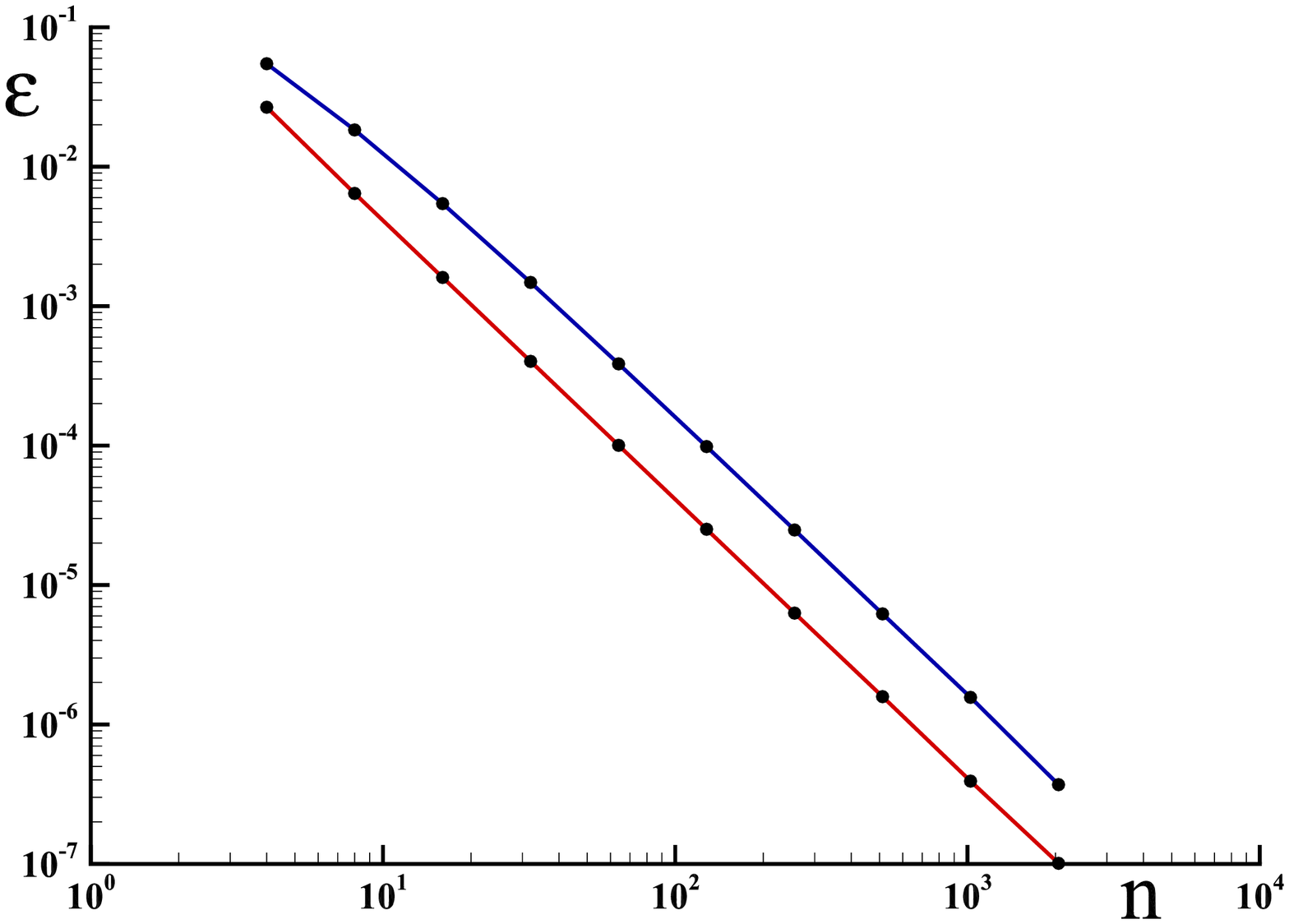}
\includegraphics[width=5.6cm,height=4.6cm]{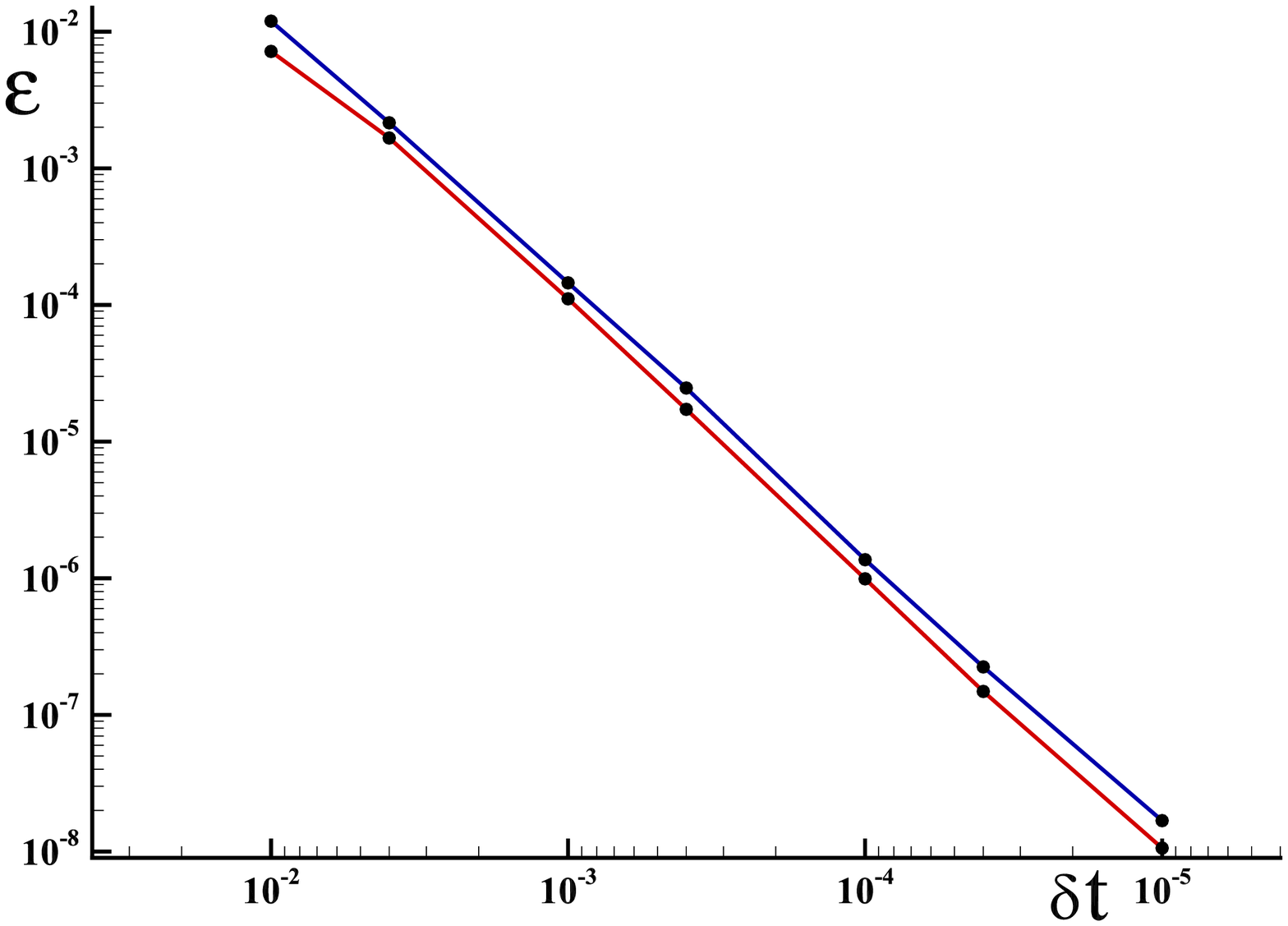}
\caption{ \it Green-Taylor vortex: convergence in space ($n$ is the number of degrees of freedom) and in time ($\delta t$ is the time step) for $L_{\infty}$ norm;  for the scalar potential $\phi$ in blue and the velocity $\mathbf V$ in red. }
\label{greenconv}
\end{center}
\end{figure}

Convergence in time at order two is obtained using a Gear scheme for the unsteady term, the linearization of the terms of inertia $| \mathbf V |^2/2$ in the form $(\mathbf V^{n} \cdot \mathbf V^{n + 1}) / 2$ and the potential upgrade by the expression $\phi^{n + 1 } = \phi^{n-1} - 2 \: \nabla \cdot \mathbf V^{n + 1}$.
Spatial convergence has been studied on a structured mesh from the stationary solution but the results are identical for a pre-defined time or in regular unstructured mesh. The number of points $ n $ per direction of space is less than or equal to $2048$. The figure (\ref{greenconv}) shows that the convergence is on order two in time and in space in norm $L_{\infty}$ on scalar potential and velocity. 

The use of the vector of Lamb $\mathbfcal {L} = \nabla \times \mathbf V  \times \mathbf V$ in continuum mechanics leads to the expression of the Navier-Stokes equation by component. However, this vector makes it possible to represent the inertial term in a planar surface $(x, y)$ for a 2D description.
It should be noted that the divergence of the Lamb vector $\mathbfcal{L}$ is not zero, even if its curl $\nabla \times \mathbfcal {L} = 0$ is really null, so that the Lamb vector derives from a scalar potential of the considered problem. In three dimensions of space, the Lamb vector is more difficult to interpret. This leads to different turbulence properties in 2D and 3D. Like all simulations performed in incompressible motions, the results of the discrete model are identical to those of the Navier-Stokes equation.

\textcolor{blue}{\subsection{Lid-driven cavity at $Re = 5000$  } }

The case of the lid-driven cavity, considered at a sufficiently large Reynolds number, is ideal for testing the legitimacy of the inertial term formulation in the equation of motion. From a general point of view, the solenoidal and irrotational parts of inertia are not easily expressed. It seems advisable to present tangible results to the numericians and CFD specialists who are attached to the inevitable Navier-Stokes equations that are widely used. They have shown their relevance in the immensity and variety of the cases modeled and simulated with them over centuries. 
Despite the different physical models and the changes brought by the equation of discrete mechanics that replaces the Navier-Stokes formulation, {\it i.e.} the treatment of pressure that is transformed into Bernoulli pressure, and the writing of the inertia term in the form $ - \nabla \times \left (\phi_i \: \mathbf n \right) + \nabla \left (\phi_i \right) $, the solutions obtained with both approaches are very close to the reference results obtained previously in the literature for the chosen Reynolds number.

The results of \cite{Bru06} for a Reynolds number of $Re = 5000$, for which the flow is stationary, are reproduced in table (\ref{cavity-5000t2}). The quantitative comparison concerns the amplitude and position of the vortices generated by the detachment of the flow on the walls of the cavity. Very good accuracy is obtained by comparing the DM formulation and the Navier-Stokes equations to reference \cite{Bru06}. Both formulations provide the same physical solution. 
A convergence study is investigated for this configuration. It shows a spatial convergence rate of $ 2 $ for velocity and pressure. The Bernoulli pressure was used to conduct the evolutions in time and the pressure itself was then extracted from it.
\begin{table}[!ht]
\begin{center}
\begin{tabular}{|c|c|c|c|c|c|c|c|c|}   \hline
 Ref.  & $\psi_{max}$ &  $x_{max}$  & $y_{max}$  & $\psi_{min}$ & $x_{min}$  &  $y_{min}$  \\ \hline  \hline
Present $256^2$  & $0.1219$  & $0.5153$  & $0.5352$  &  $-3.086 \: 10^{-3}$ & $0.8040$  &  $0.07310$  \\ \hline
Bruneau al. $2048^2$& $0.12197$ & $0.515465$  & $0.53516$  &  $-3.0706 \: 10^{-3}$ & $0.80566$  &  $0.073242$  \\ \hline
\end{tabular}
\caption{\it Comparison of the results obtained in mechanics of continuous media \cite{Bru06} and those resulting from the DM formulation presented at $Re =  5000$ for a Chebyshev mesh with $256^2$ cells. } 
\label{cavity-5000t2}
\end{center}
\end{table}

Discrete Mechanics does not bring into question the results obtained with the Navier-Stokes equation. As previously observed, the same solutions are obtained at least with a convergence order and an accuracy that are almost identical to computer errors. However, the most amazing thing is that continuous and discrete models differ on many points. 
\begin{figure}[!ht]
\begin{center}
\includegraphics[width=4.5cm,height=4.5cm]{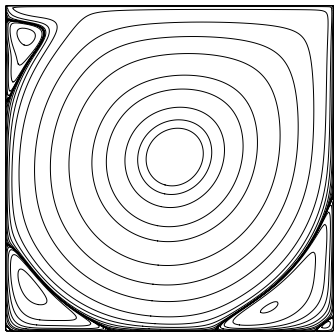}
\caption{ \it Lid driven cavity at $Re = 5000$ with Cartesian mesh (mesh-cart-7.typ2); $\psi_{max} = 0.1211$ at $ x_{max} = 0.5152$ and $ y_{max} = 0.5353$ and $\psi_{min} = - 0.003138$ at $ x_{min} = 0.8024$ and $ y_{min} = 0.07180$.  }
\label{cavite-5000}
\end{center}
\end{figure}

One of these fundamental differences is the use of mass conservation that is always associated with the Navier-Stokes equation. On the contrary, the discrete mechanics equation does not use it explicitly and the mass is always strictly conserved for compressible or incompressible movements: the DM formulation behaves as an autonomous equation that does not require any additional constitutive law. In fact, mass conservation is implicitly integrated into the equation of motion through the term $dt \: c_l^2 \: \nabla \cdot \mathbf V$ \cite{Cal11}.
The stream lines shown in figure (\ref{cavite-5000}) are obtained directly from the vector potential $\bm \psi = - \nu \: \nabla \times \mathbf V$ projected on the surface, $\psi = \bm \psi \cdot \mathbf n$.

The lid driven cavity example is a classical flow that shows all the interest of the discrete formulation compared to a continuum type approach. First of all, the solutions with constant properties are strictly the same, whatever the problem dealt with, from the analytical solutions of the Navier-Stokes equations to complex flows whose solutions are obtained numerically. The interest resides in the physical understanding that can be exhibited from the scalar and vector potentials that make the equation of discrete mechanics a true extractor of Hodge-Helmholtz components of acceleration.

\textcolor{blue}{\subsection{Droplet acted upon by gravity} }

To describe the wetting of a realistic surface, we must overcome a range of obstacles associated with defining, characterizing, and modeling the capillary effects in the presence of three separate media -- usually a gas, a liquid, and a solid -- which meet at the so-called triple line. Flows involving capillary effects depend on two characteristic properties: the surface tension $\sigma_{ij}$ between the $i$-th and $j$-th media, and the local curvature of the free surface $\Sigma$ of the immiscible media. Do we really need to introduce another physical parameter to describe the concept of wetting? For the past few decades, the proposed solution has been to observe the contact angle $\theta$ between two of the media, typically a liquid surface and a solid substrate. This contact angle represents a static measurement that is deduced from an array of theoretical principles and experimental methodologies. Much has been written about contact angles in the literature, but since the perspective developed here diverges significantly from past work, we shall focus on a presentation that aligns with the discrete approach. When the interface is in motion, past approaches have proceeded by introducing a time-dependent contact angle known as the dynamic contact angle; various observation-inspired laws have been formulated to describe the evolution of angle $\theta$ over time under certain specific circumstances.

We shall tackle the problem at a global level from the very start -- since we ultimately want to know the acceleration $\bm \gamma$ of the fluid on the edge $\Gamma$, it must be possible to phrase any actions on this edge as a sum of contributions. First of all, there must be inertial and viscous accelerations, a gravitational acceleration, and finally a capillary acceleration $\gamma_c$. The motion of the triple line depends on all of these effects; it doesn't make sense to define a dynamic contact angle that depends exclusively on time {\it a priori}. The model proposed by discrete mechanics is based on the following observation: if the interface $\Sigma$ with the substrate is allowed to move freely, it will naturally attempt to recover the equilibrium position defined by the static contact angle $\theta$. The dynamics themselves are governed by the acceleration $\bm \gamma$. The physical behavior of the interface in the neighborhood of the triple line is determined by the need to establish a mechanical equilibrium from the accelerations of each pair of media. The curvature of the interface in this region is in fact the most effective parameter for describing the tendency of the system to return to equilibrium. If the so-called contact curvature $\kappa_c$ \cite{Cal19a} is known, then, although the values of the instantaneous curvature $\kappa$ depend on each of the various effects, they will tend toward $\kappa_c$ at mechanical equilibrium. For example, in the absence of gravity, a droplet of liquid on a non-deformable solid substrate will tend toward having the curvature $\kappa_c$ that corresponds to the static contact angle $\theta$; at equilibrium, the entire interface will have curvature $\kappa_c$. The relation between the contact curvature $\kappa_c$ and the contact angle $\theta$ is established by considering the geometry of the interface. The key advantage of the proposed model is that it avoids introducing a new parameter, since the curvature is already part of the model of $\bm \gamma_c$. We can simply require the curvature to be equal to $\kappa_c$ on the triple line and allow the dynamics of the system to be governed naturally by the equations of motion themselves.

The spherical-cap shape is modeled directly and the curvature on the primal topology can be calculated by a formula from differential geometry. In the case of a spherical cap, the curvature is exact $(\kappa = 1/R)$ up to the machine error. Since the contact curvature is $\kappa_c = \kappa$, the velocity is zero, but the capillary pressure is equal to $p_c = \gamma \: \kappa = \sigma \: \kappa / \rho_v $, as expected. We will consider the case of the unsteady evolution of a droplet on a planar surface later.
Any droplet of liquid on a surface that does not satisfy the equilibrium
\begin{eqnarray}
\displaystyle{  - \nabla \phi^o  + \nabla \left( \sigma \: \kappa \: \xi \right) = 0 }  
\label{mouillec}
\end{eqnarray}
will necessarily generate motion and oscillations due to the inertial effects that lead it toward the equilibrium state with potential $\phi_c =  \sigma \: \kappa$. 

In the presence of gravity, but without hysteresis or inertial effects, the velocity of the droplet on a surface inclined at an angle of $\varphi$ is constant and the acceleration is zero; the equation governing the mechanical equilibrium in this case may be stated as follows:
\begin{eqnarray}
\displaystyle{  - \nabla \phi^o  + \nabla \left( \sigma \: \kappa \: \xi \right) +  \mathbf g \: \sin{\varphi}  = 0. }\label{mouilled}
\end{eqnarray}

Gravity introduces an asymmetry between the advancing contact angle and the receding contact angle. As a result, the droplet no longer has the shape of a spherical cap or a circular arc. The solution can only be found by solving aquation (\ref{discrete}) with the source term corresponding capillary effects.
The notions of contact angle and contact curvature are of course linked -- bijectively, in the case of a static equilibrium without external forces.
\begin{figure}[!ht]
\begin{center}
\centering\includegraphics[width=6.5cm]{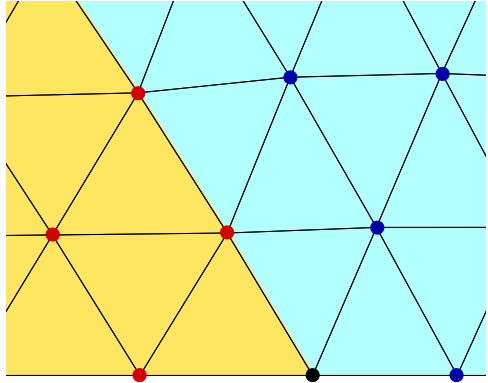}
\caption{\it Model of the wetting of a droplet on a planar surface. The points \large $\bullet$ are the first fluid and at the interface, where $\xi =$ {\em 1} and $\kappa^* = \kappa$, the points \textcolor{blue}{\large $\bullet$} are in the second fluid, where $\xi =$ {\em 0}, and the point \large $\bullet$ is on the triple line, where $\xi =$ {\em 1} and $\kappa_{\hbox{\scriptsize c}} = \kappa$.}
\label{calotte-equil}
\end{center}
\end{figure}

From a physical point of view, the contact curvature is a local and intrinsic property of the surface; this is the maximum curvature that the interface can sustain before a flow is generated to find another equilibrium position. For example, in the case of the lotus effect, the mean surface can be viewed as non-wetting, but this property arises from the high curvature of the interface around the hairs of the leaf. Any gravitationally induced effects are compensated by variations in the curvature, which is essentially constant at a sufficient distance away from the contact points. In the general case, it seems impossible to predict the at rest contact angle in the presence of forces such as gravity.

The static and dynamic equilibria themselves are physically complex and it is difficult to give an answer that is both simple and general. The strategy adopted here is to construct a set of equations of motion that can account for any potential capillary effects, then apply this model to perform direct simulations. The term $\nabla ( \sigma \: \kappa \: \xi )$ seems to incorporate all of the desired effects. The examples given below help to justify this perspective.

By using this concept of contact curvature, we can guarantee that the local curvature in the neighborhood of the triple line is equal to $\kappa_c$ at the static equilibrium. Without gravity, in the case of perfect wetting, $\kappa_c$ is equal to zero. For a perfectly non-wetting surface, the contact curvature must be smaller or equal to the inverse of the radius of the sphere that would no longer be in contact with the surface. Again, the desired physical phenomena can only be reproduced if the equations of motion themselves are representative. The contact curvature directly describes the equilibrium or disequilibrium in terms of the resultant of the accelerations on the triple line. 
Phenomena involving hysteresis are not addressed here, but they can be approached with the same formalism by replacing the advancing and receding contact angles with two distinct contact curvatures.

The triple line represented by the black point in figure (\ref{calotte-equil}) belongs to both the interface and the solid wall. We can assign a special value of the curvature to this point, the contact curvature $\kappa_c$. If this value is the same as the rest of the interface, then, like for the simulation with a structured mesh, the velocity is zero and the overpressure in the droplet is therefore equal to $p_c = \gamma \: \kappa$. If $\kappa_c \ne \kappa$, the forces are no longer balanced on the triple line, and the resultant generates motion that spreads to the rest of the droplet by continuity.
In conclusion, whether or not the interface aligns with the mesh, the equilibrium of droplet is guaranteed whenever $\kappa_c = \kappa$ in the absence of gravity, and the contact angle $\theta$ can also be used to describe the static equilibrium. If gravity or some other force is present, the contact angle relation is no longer satisfied, but the concept of contact curvature remains applicable at static equilibrium.

\begin{figure}[!ht]
\begin{center}
\includegraphics[width=10.5cm,height=3.5cm]{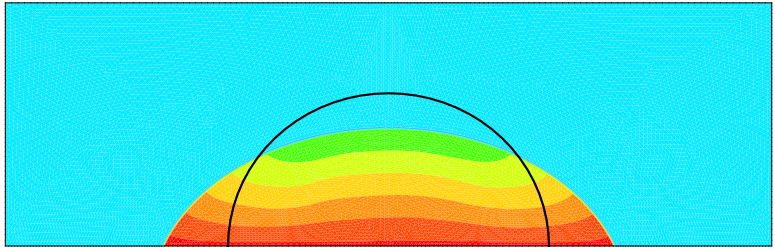}
\caption{\it Equilibrium of a droplet of radius R $=$ {\em 6.25 $\cdot$ 10}$^{-{\hbox{\scriptsize\em  4}}}$ under gravity g $=$ -{\em 100}  m s$^{-{\hbox{\scriptsize\em  2}}}$ from an initial state where the droplet is hemispherical and has zero velocity. The maximum pressure variation in the pressure field is $\Delta$p $=$ {\em 188.9} Pa, and the curvature $\sigma$ takes values in {\em [835, 1574]}. Conservation of mass is satisfied up to the machine error.}
\label{calotte-grav}
\end{center}
\end{figure}

When gravity is present, the static equilibrium of a rest state is more delicate to describe. Exact solutions are much less common; in the general case, a solution can only be given by solving the equations of motion (\ref{discrete}) numerically.

Consider a semicircular droplet initially at equilibrium on a surface with a contact angle of $\theta = 90$Ю At the initial moment, gravity is increased from $0$ to $g= - 100 \: m \: s^{-2}$. The weight of the droplet causes it to flatten out and significantly changes its shape. The curvature, initially equal to $\kappa = 1600$, becomes variable along the interface, but the maximum curvature gradually tends toward the imposed value of $\kappa_c = 1600$ over time.

Figure (\ref{calotte-grav}) shows a rest state very close to the static equilibrium. Gravity exerts a noticeable effect on the pressure field.
Unlike the contact angle $\theta$, the concept of contact curvature $\kappa_c$ is extremely robust. It simply describes the action of any effects applied to the triple line. In the description of a discrete medium, any forces are replaced by accelerations, which provides a number of advantages. Furthermore, we don't need to introduce any other information than is already present in the discrete equations of motion. The concept of curvature, which is already defined at any interface to describe the capillary effects, can also be used to describe phenomena involving the triple line.

The system (\ref{discrete}) does not require any boundary conditions, neither on a vector variable such as the velocity component $\mathbf V$ nor a scalar variable. Every constraint is incorporated implicitly \cite{Cal19a} in the same way as partial wetting.

\textcolor{blue}{\subsection{A monolithic approach of Fluid-Structure Interaction} }

Even if the rheology of the medium is more complex, e.g. viscoelastic fluids, non-linear viscosity laws, viscoplastic fluids, time-dependent properties, and so on, we should still be able to represent its behavior under various types of applied stress.
In some cases, the shear-rotation stresses may only be partially accumulated. We can describe viscoelastic behavior by weighting the accumulation term of $\bm \psi^o$ by an accumulation factor $0 \leq \alpha_t \leq 1$. Fluids with thresholds can also easily be represented by specifying a value of $\bm \psi^o = \bm \psi_c$ below which the medium behaves like an elastic solid. Many of the difficulties that are typically encountered in rheologies with non-linear viscosities are no longer an issue with this model.

In discrete mechanics, the concepts of viscosity and shear-rotation are exclusively associated with the faces of the primal topology, where the stress may be expressed in the form $\nu \: \nabla \times \mathbf V$ in fluids and $dt \: \nu \: \nabla \times \mathbf V$ in solids.
As an example, let us examine the interaction of an incompressible viscous Newtonian fluid and a neo-Hookean elastic solid. The stress tensor of an incompressible isotropic hyperelastic material is as follows in the neo-Hookean model:
\begin{eqnarray}
\displaystyle{ \bm \sigma_s = - p \: \mathbf I + \mu_s \: \mathbf B,  }
\label{hook}
\end{eqnarray}
where $\mathbf B = \mathbf F \: \mathbf F^t$ is the left Cauchy-Green deformation tensor.
In two spatial dimensions, the Cayley-Hamilton theorem can be used to show that the Mooney-Rivlin model of a hyperelastic material is equivalent to the neo-Hookean model.
\begin{figure}[!ht]
\begin{center}
\includegraphics[width=6.5cm]{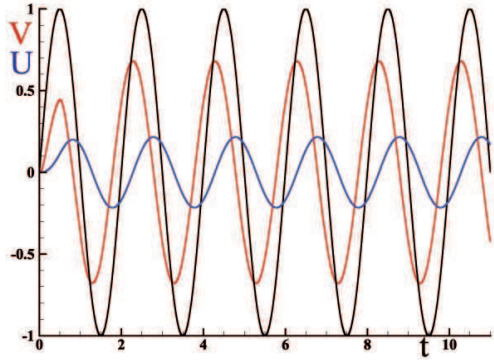}
\caption{\it Study of a periodic fluid-structure interaction between a viscous fluid and an elastic solid; the viscosity of the fluid is $\nu = 1 \: m^2 \: s^{-1}$ and the shear modulus of the solid is $\nu = 4 \: m^2 \: s^{-1}$. The velocity of the fluid at the upper wall is shown in black, the velocity of the interface $\Sigma$ is shown in red, and the displacement over time of the solid \textbf{\em U} at the interface is shown in blue.}
\label{sugiyamatime}
\end{center}
\end{figure}

We shall study a problem that was published by Sugiyama in 2011 \cite{Sug11}. Consider an elastic band with an applied shear stress generated by the periodic flow of an incompressible Newtonian fluid. The flow is laminar and periodic in $x$. Given that there are no compression terms, we can solve the problem in one spatial dimension along the $y$-axis for $y \in [0, 1]$. Suppose that the upper interface follows the periodic motion $V(t) = V_0 \: \sin ( \omega \: t )$, where $V_0 = 1$ and $\omega =  \pi$. The velocity of the lower surface is kept at zero. The solid occupies the lower part of the domain, and the fluid occupies the upper part of the domain; the position of the interface is $y=1/2$. The theoretical solution found by Sugiyama was obtained by separating the spatial variable $y$ from the time variable $t$; a homogeneous solution is found by considering a basis of Fourier functions on the interval $y \in [0,1]$ and exponential functions on the time interval in each of the fluid and solid domains separately. The sequence of Fourier coefficients can be determined from the coupling at the interface by requiring the velocity and the stress to be continuous.
We can find a solution $V(y,t)$ directly from the equations of discrete mechanics (\ref{discrete}) simply by imposing the relevant conditions at $y=0$ and $y=1$. The coupling conditions at the interface, namely the continuity of the velocity and the stress, are implicitly guaranteed to hold by the dual curl operator. The notion of a 2D or 3D space does not exist in discrete mechanics. Instead, the operators define the orientations of the normal and tangent directions within a three-dimensional space. Despite this, the hypotheses of this example enable us to solve along a single spatial dimension. The time step is chosen to be $\delta t = 10^{-4}$ to ensure good overall levels of accuracy; by comparing against the theoretical analytic solution, it can be shown that the numerical solution is second-order in space and time (this result is included in textbook \cite{Cal19a}).

Figure (\ref{sugiyamatime}) plots the velocity and the displacement of the interface $\Sigma$ over time. The velocity of the upper wall is also shown.
The solution establishes itself very quickly. After just a few periods, the velocity becomes fully periodic. The velocity profiles are shown until $t=10 \: s$. The displacement of the solid over time may be deduced from the relation $\mathbf U = \mathbf U^o + \mathbf V \: dt$, where $dt$ represents both the differential element and the time increment $\delta t = dt$.
Note that the displacement is strongly out of phase with the velocity of the interface.

\begin{figure}[!ht]
\begin{center}
\includegraphics[width=3.7cm,height=3.7cm]{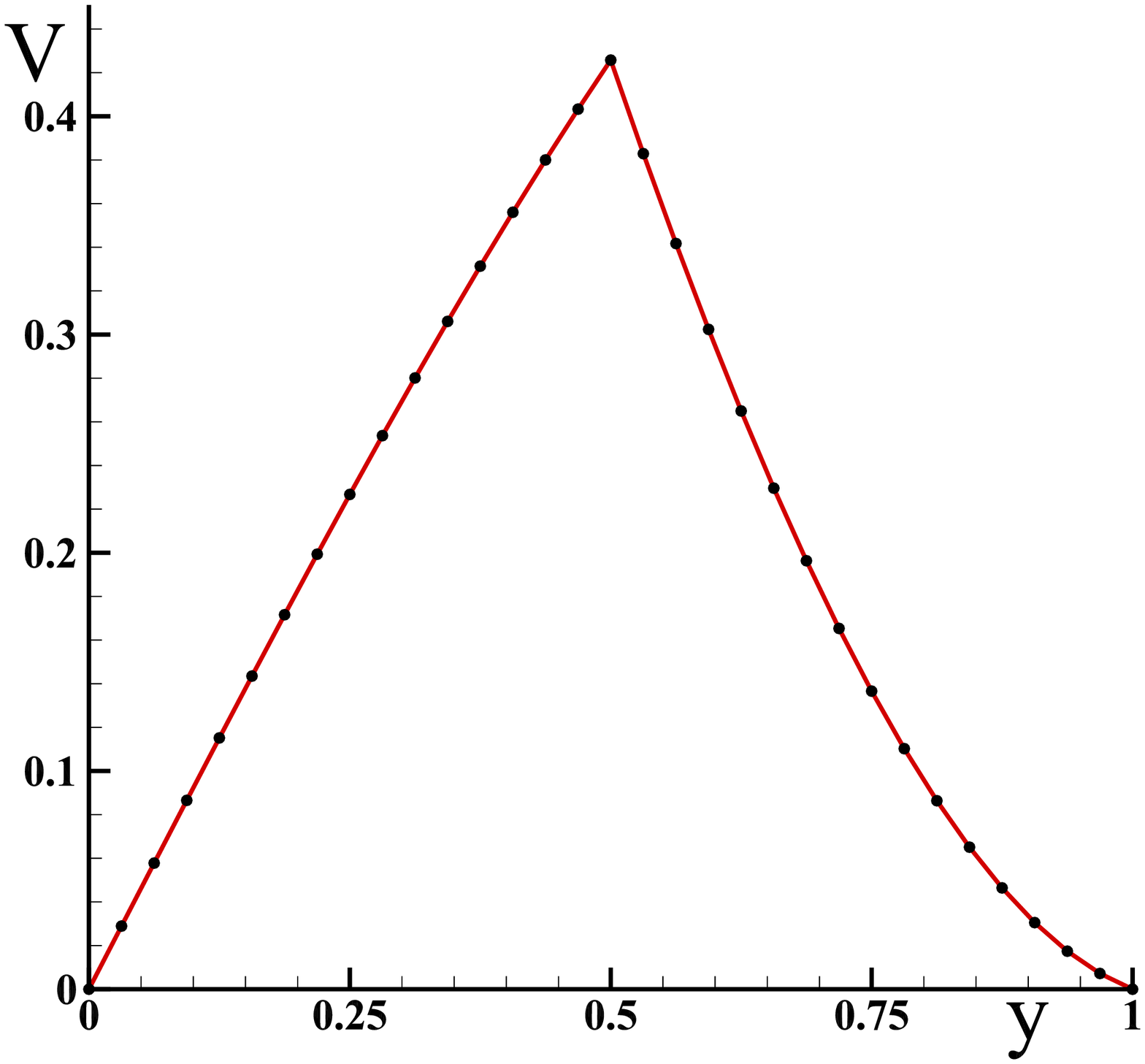}
\includegraphics[width=3.7cm,height=3.7cm]{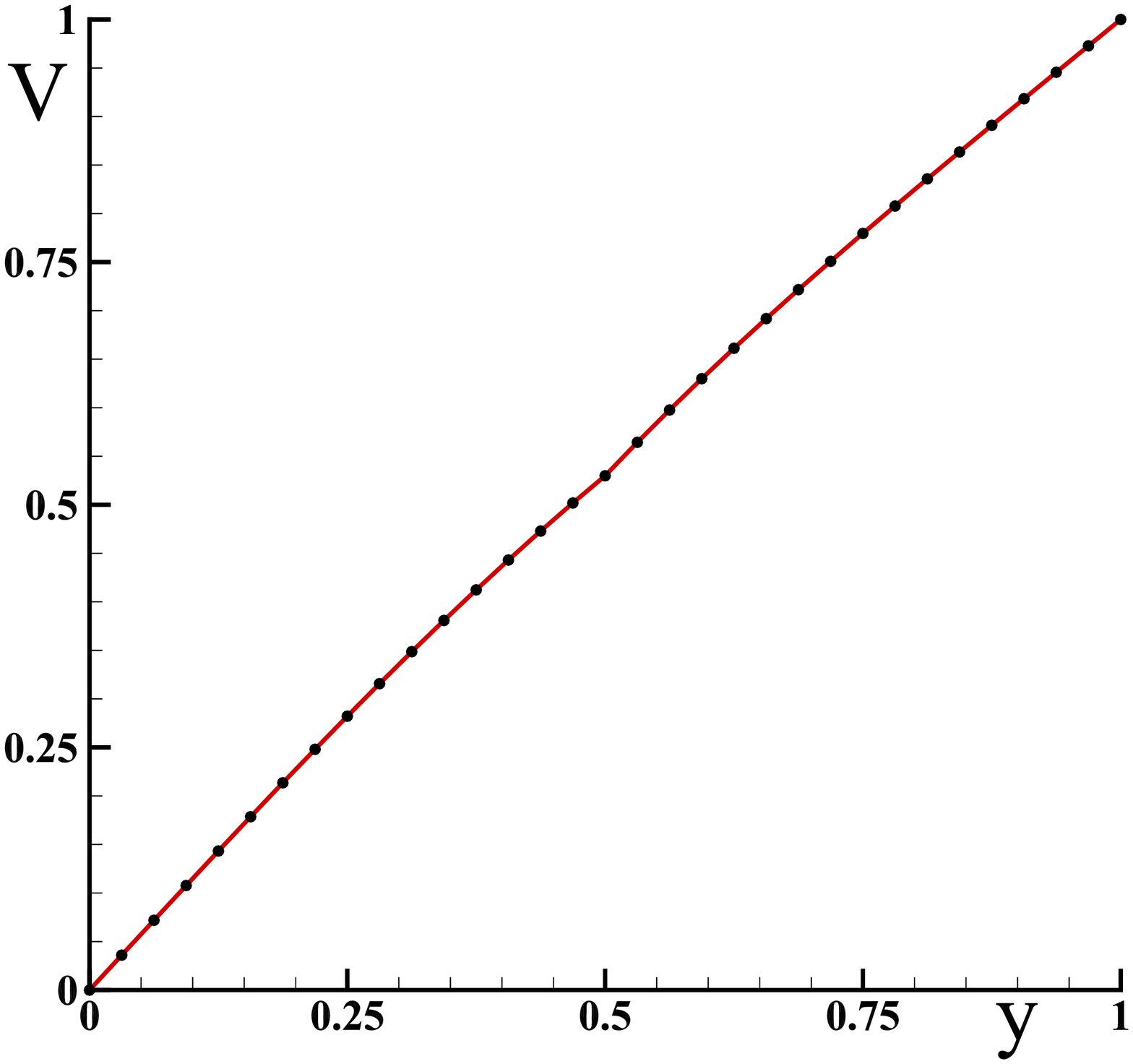}
\includegraphics[width=3.7cm,height=3.7cm]{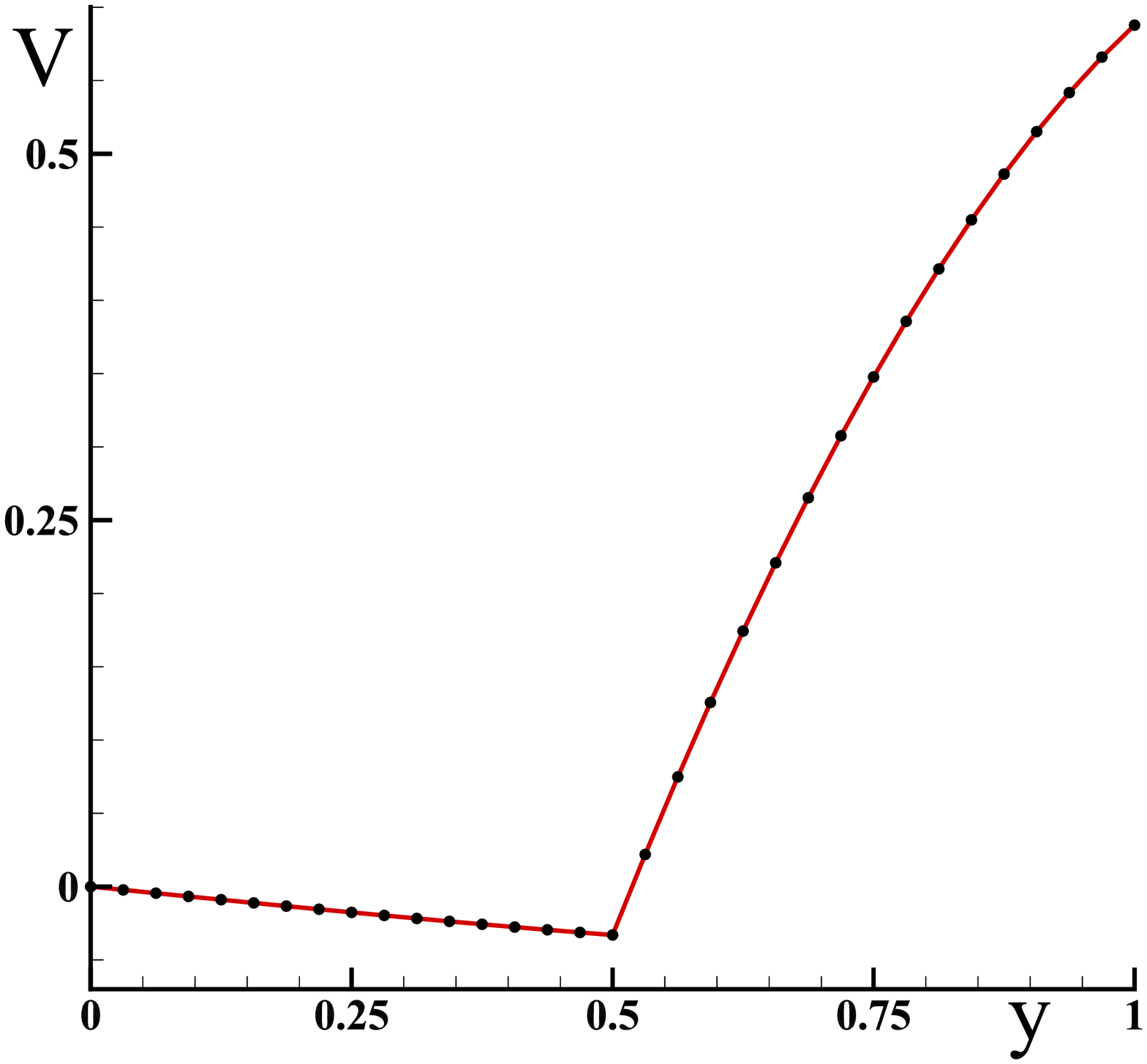}
\caption{\it Study of the fluid-structure interaction between a viscous fluid and an elastic solid with a periodic evolution. The viscosity of the fluid is equal to $\nu = 1 \: m^2 \: s^{-1}$ and the equivalent solid shear modulus is equal to $dt \: c_t^2 = 4 \: m^2 \: s^{-1}$. Velocity profiles based on $y$ for times $t = 10 \: s$, $t = 10.5 \: s$, $t = 10.8 \: s$ are plotted. The solid lines show the theoretical solution, while the dots correspond to a spatial approximation of a $32$ mesh for $y \in [0,1]$.}
\label{sugiyamavts}
\end{center}
\end{figure}

A selection of the velocity profiles in the $y$-direction are shown in figure (\ref{sugiyamavts}) once the periodic regime is fully established. The results converge to second order in space and time. Given the absolute accuracy (of the order of $10^{-4} \: s$) obtained using a coarse mesh $(n = 32)$, we can conclude that there is no observable error between the theoretical solution and the numerical solution.

One advantage of the fluid-structure interaction for a neo-Hookean model described by Sugiyama is that it has a theoretical solution. This allows us to compare the numerical solutions that we obtain more precisely, but also allows us to develop new concepts, as we did for discrete mechanics in this section. Sugiyama obtained a first-order error in the $L_2$ and $L_{\infty}$ norms, whereas the model (\ref{discrete}) achieves second-order results with much lower absolute errors.
This improvement is ultimately attributable to the separation of the properties at the interface, as well as the fact that no interpolation is performed, despite a fully monolithic and implicit treatment of the fluid-solid coupling.

Fluid-structure interactions on 2D or 3D geometries with a moving interface can of course also be solved using the system (\ref{discrete}). However, without an analytic solution for comparison, there is little benefit in doing so, since the errors of the various methodologies accumulate over each step of the process. 
Other more complex constitutive laws can also be modeled. However, although specific research in this domain is interesting in its own right, this does not offer any additional validation of the discrete model. The complete separation of the laws of motion and the constitutive and state laws in discrete mechanics will likely enable us to apply any kind of law to problems such as this one.

\textcolor{blue}{\section{Conclusions} }

The original form (\ref {discrete}) of the discrete motion equation corresponds to a physical model different from the one leading to the Navier-Stokes equation. It is based on the Hodge-Helmholtz decomposition of the acceleration into a component with divergence free and another irrotational.
The velocity, the variable of the vector equation, makes it possible to upgrade the scalar and vector potentials which physically represent respectively the compressive and rotational-shear stress.
 This decomposition was possible considering that the physical properties are not differentiable quantities in the sense of the continuum. Considering these properties piecewise constant still allows any problem of fluid mechanics or unrestrained solids to be addressed. Two-phase flows with highly variable properties, the elasticity associated with complex constitutive laws, the fluid-structure interaction are accessible by solving equation (\ref {discrete}).

The declination of the formulation $ (\mathbf V, \phi, \bm \psi) $ in the form of a splitting has conducted to an original projection method. This projection method derived from the discrete formulation is described as exact insofar as it results from the application of the divergence operator to the equation of motion to give a Poisson equation with constant coefficients (\ref {divinert}) without any hypothesis or simplification. It has the same disadvantage as the other projection methods, the existence of a boundary layer near the walls, but it also has several decisive advantages over the other methodologies like for multiphase flows with high density ratios.
 This technique is order two in time like the formulation $ (\mathbf V, \phi, \bm \psi) $.

\vspace{7.mm}

\bibliographystyle{elsarticle-num} 

\bibliography{database}

\end{document}